%% file: hysea2014-paper.tex
\documentclass[12pt,final]{elsarticle}
\usepackage{amsmath,amssymb,color}
\usepackage{natbib}
\definecolor{ze}{RGB}{0,64,0}
\usepackage{hyperref}
  \usepackage{paralist}
\usepackage{relsize}
\usepackage{graphicx}
\usepackage{tikz}
\usepackage{eso-pic}
\usepackage{pdfpages}
\usepackage{graphicx,eso-pic,lipsum}
\usepackage{bm}
\usepackage{bbm}
\usepackage[notcite,notref]{showkeys}
\usepackage[normalem]{ulem}

\definecolor{gray}{rgb}{0.9,0.9,0.9}

\newcommand\bfh{\mathbf h}

\newcommand\bfn{\mathbf n}

\usepackage{mathrsfs}
\renewcommand\mathcal{\mathscr}

\newcommand\bfu{\mathbf u}

\def\bfS{{\mathbf S}}

\def\bf0{{\mathbf 0}}
\def\bfn{{\mathbf n}}

\def\tr{{\mbox{tr}}}

\journal{...}

\begin{document}
\begin{frontmatter}
\title{Stress effects on the kinetics of adsorption in a spherical particle: an analytical model}

 \author{Fernando P. Duda}
 \ead{duda@mecanica.coppe.ufrj.br}
\address{Programa de Engenharia Mec\^anica, COPPE/UFRJ, Rio de Janeiro, Brazil}
\author{G. Tomassetti
}
\ead{tomassetti@ing.uniroma2.it}
 \address{Universit\`a di Roma ``Tor Vergata'', Dipartimento di Ingegneria Civile e Ingegneria Informatica, Via Politecnico 1, 00169, Italy.} 

 \begin{abstract}
 We consider a two-phase elastic solid subject to diffusion-induced phase transformation by an interstitial species provided by a reservoir.  We derive a simple analytical model to quantify the effect of misfit strain on the kinetic of phase transformation and to calculate the amplitude of the well-know hysteresis cycle observed when a sequence of forward and reverse phase transformations takes place.
 \end{abstract}


 

\end{frontmatter}

\allowdisplaybreaks




\section{Introduction}
Metallic nanoparticles are characterized by fast hydrogenation and dehydrogenation kinetics and hence are of particular interest for hydrogen-storage applications. A source of complication in the understanding of the adsorption/release kinetics in these devices is the stress accompanying hydride formation due to the misfit between the metal and the hydride lattice structures.

It is well known that the elastic misfits associated to phase transformation give rise to a stress field that may affect considerably phase equilibria in multiphase elastic solids \cite{fratzl1999modeling,larche1985overview}. In this respect, of particular importance is the analysis carried out by Schwarz and Kachaturyan \cite{SchwaK1995Thermodynamics}, who examined a two-phase solid solvent in contact with a reservoir providing solute interstitial atoms. Their analysis shows that transformation-induced strain makes it impossible for the two phases to coexist at equilibrium and that, moreover, it is responsible for the hysteresis loop observed in a cyclic adsorption-desorption processes; however, it does not address the issue of how misfit strain may affect phase-transformation kinetics.

In order to investigate this issue, in this paper we address the kinetics of phase-transformation in a spherical particle in contact with a reservoir of interstitial atoms at prescribed chemical potential $\mu_R$, as shown in Fig. 1 below.
\begin{figure}[h]
\begin{center}
\def\svgwidth{0.3\linewidth}
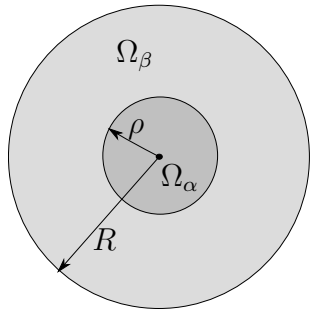
\footnotesize\caption{Spherical specimen. }
\end{center}  
\end{figure}
General multi-field theories have been devised which describe the concomitant processes taking place in the solid, such as phase transformation, deformation, and diffusion \cite{BonetFL2007AMO,VoskuP2013Phase,RoubiT2014DCDS}. In order to arrive at an analytically-tractable model, we simplify the problem by assuming that there exists a \emph{sharp interface} which separates the body in two phases:
\begin{itemize} 
\item a low-concentration $\alpha$ phase, occupying the region $\Omega_\alpha$ and
\item a high-concentration $\beta$ phase, which occupies the region $\Omega_\beta$.
\end{itemize}
By changing the pressure (and hence the chemical potential) of the reservoir, adsorption/desorption of atom takes places, and, in order to accommodate the increased/decreased amount of diffusant, the phase boundary, assumed to be \emph{sharp} and \emph{coherent}, is set in motion. 

We base our model on the continuum theory of elastic solids coupled with species diffusion developed in \cite{fried1999coherent}. In particular, we use the notion of configurational force as a tool to describe the evolution of the material structure of a body --- in this case, the phase interface. 

A crucial assumption is that chemical potential does not depart from the reference chemical potential $\mu_0$, where the two phases would be at equilibrium in the absence of misfit strain. This value of chemical potential represents the equilibrium state usually considered in thermodynamics, for which surface and elastic effects play no role. This fact provides us with motivation to assume that the Gibbs energy depends linearly on the difference between the local value of chemical potential and the reference one. A consequence of this fact is that concentration is prescribed on each phase. 

We find, as a result, that the evolution of the phase boundary is governed by a differential equation whose solution can be explicitly worked out, and provides us with a precise assessment of the characteristic time of adorption and desorption. Our framework is fairly general, and may be adapted to several applications of technological relevance. In particular, it may be used to characterize the kinetics of hydrogen adsorption in spherical nanoparticles.

A similar problem has already been considered in \cite{MishiB2010AM}, where hydride formation and dissolution was studied under the assumption of small departure of chemical potential from a reference value. The study  in  \cite{MishiB2010AM} is limited to nucleation. Here, we consider both phase nucleation and phase growth. Effects of lattice strains on kinetics of hydration/dehydration have also been considered in \cite{ZhdanK2009JPCC,Zhdan2010CPL,LanghZZK2010CPL}, under different modeling assumptions.

It is also worth noticing that our model is intimately connected with that describing a coherent precipitate from a solid solution, a problem treated by Voorhees and coworkers in \cite{LaraiJV1988JMR,LaralJV1989SM} (see also the review \cite{fratzl1999modeling}, as well as the paper \cite{LeoS1989AM}). In their investigation of the role played by elastic effects on the  growth  kinetics, they consider that the precipitate grows in an infinity matrix.  The boundary conditions for the diffusion equation in the matrix, which is an equation for the composition field,   stipulated two values for the composition,  a far field one and a interfacial one, the latter obtained from the Gibbs-Thompson condition.

The aforementioned problem is amaneble to analytical treatment under certain simplifying assumptions, including that the precipitate is spherical and that diffusion is fast in comparison to the the  interface motion.  Thus, by considerind small departures about the incoherent equilibrium (no elastic and surface effects), a kinetic equation for the rate
of growth of a spherical inclusion of radius $\rho$ in an infinite elastically isotropic matrix is obtained (see equations (75) to (77) of \cite{LeoS1989AM}). This result can be obtained from the result of our paper by assuming that $R$ goes to infinity, neglecting capilarity effects, and relating $\mu_R$ with the far-field composition.

\section{Setting the stage}
In this section we assemble the ingredients of the continuum theory of elastic solids undergoing diffusion and phase transformation expounded in \cite{fried1999coherent}, to which we refer those readers who may want additional details. 

We consider the three-dimensional elastic body $\Omega$ pictured in Fig. 1. The body has the shape of a sphere of radius $R$, and is partitioned into two time-dependent regions: the region ${\Omega}_\alpha(t)$, occupied by the $\alpha$ phase, and the region ${\Omega}_\beta(t)$, occupied by $\beta$ phases. We denote by $\varrho(t)$ the radius of $\Omega_\alpha(t)$ and by by ${\cal S}(t)$ its boundary, namely, the evolving interface separating the two phases. We assume that such spherical symmetry be preserved during the evolution process. The fields that are relevant to our description are: 
\begin{itemize}
\item concentration $c$;
\item chemical potential $\mu$;
\item stress $\mathbf S$;
\item displacement $\mathbf u$;
\item linear strain $\bm{\varepsilon}$;
\item flux of diffusant $\mathbf h$;
\item Helmoholtz free energy per unit volume $\psi$;
\item Gibbs free energy per unit volume $\omega$.
\end{itemize}
Each field depends on the typical point $x\in\Omega$ and on time $t$. Linear strain and displacement are related by the \emph{compatibility condition}:\footnote{Here ${\rm sym}$ denotes the symmetric part of a tensor. In components, \eqref{eq:32} reads: $\varepsilon_{ij}=\frac 12 (u_{i,j}+u_{j,i})$, with a comma denoting partial derivative.}
\begin{equation}\label{eq:32}
  \bm\varepsilon=\textrm{sym}\nabla\mathbf u=:\bm\varepsilon(\mathbf u).
\end{equation}
Stress and flux of diffusant must comply with the balance of standard forces and mass balance of diffusant, namely,\footnote{
Using Einstein's convention for repeated indices, the balance equations \eqref{eq:3} are rendered componentwise as follows:
\begin{equation*}
\sigma_{ij,j}=0,\qquad h_{j,j}=0.
\end{equation*}
}
\begin{equation}\label{eq:3}
  {\rm div}\bm\sigma=0,\qquad{\rm div}\mathbf h=0.
\end{equation}
We enforce the above-mentioned balance equations away from the interface $\mathcal S$. The corresponding balance statements at the interface take the form of \emph{jump conditions}:
\begin{equation}
\label{fbsci}
[\![\bm\sigma]\!]\bfn=\bf0, \quad [\![c]\!]V-[\![\bfh]\!]\cdot\bfn=0.
\end{equation}
Here $\mathbf n$ is the unit vector field pointing in the radial direction; $V$ is the velocity of the interface in the direction of $\mathbf n$; double square brackets enclosing a field denote the jump of that field at the interface. The first of \eqref{fbsci} encodes the requirement that the normal traction be continuous across the interface; the second, that there be no production/depletion of diffusant at the interface.

In addition to \eqref{fbsci}, we enforce at the interface the following conditions:
\begin{equation}
\label{eq1}
[\![\mathbf u]\!]=\bf0, \quad [\![\mu]\!]=0.
\end{equation}
The jump conditions \eqref{eq1} state that both \emph{displacement and chemical potential are continuous at the interface}. Such requirements encode our expectation that the interface be \emph{coherent}, that is to say, that the displacement field does not jump (and hence no crack appears) at the interface. 

Continuity of chemical potential at the interface is known as \emph{assumption of local chemical equilibrium}. The conditions \eqref{eq1} at the interface are accompanied by:\footnote{Operating pressures for hydrogen-storage in metallic hydrides are of the order of $10^6-10^7$Pa. Such pressures, by themselves, produce modest strains and can be neglected.}
\begin{equation}
  \bm\sigma\mathbf n=0,\qquad\mu=\mu_R
\end{equation}
at the boundary of the specimen.

The Gibbs free energy per unit volume is determined by specifying two functions $\widehat\omega_\alpha(\mu,\bm\varepsilon)$ and $\widehat\omega_\beta(\mu,\bm\varepsilon)$ such that the constitutive equation:\footnote{Equation \eqref{eq:34} is read as follows: given a point $x\in\Omega$ and a time $t$ such that $x$ is not part of $\mathscr S(t)$,: $\omega(x,t)=\widehat\omega_\alpha(\mu(x,t),\bm\varepsilon(x,t))$ if $x\in\Omega_\alpha(t)$, and $\omega(x,t)=\widehat\omega_\beta(\mu(x,t),\bm\varepsilon(x,t))$ if $x\in\Omega_\beta(t)$.}
\begin{equation}\label{eq:34}
  \omega=\widehat\omega_\phi(\mu,\bm{\varepsilon})\textrm{ in }\Omega_\phi\quad\textrm{for}\quad \phi=\alpha,\beta,
\end{equation}
holds at all times. Once the constitutive mapping delivering Gibbs energy has been prescribed, concentration and stress in phase $\Omega_\phi$ are given by, respectively,
\begin{equation}
  \label{eq:54}
   c=-\frac{\partial\widehat{\omega}_\phi(\mu,\bm{\varepsilon})}{\partial\mu}\quad \textrm{and}\quad \bm{\sigma}=\frac{\partial\widehat{\omega}_\phi(\mu,\bm{\varepsilon})}{\partial\bm{\varepsilon}}.
\end{equation}
As to the flux of diffusant, we assume it to have the form:
\begin{equation}\label{eq:6}
  \mathbf h=-m_\phi\nabla\mu,
\end{equation}
where $m_\phi$ denotes the \emph{mobility} of the diffusant in phase $\phi=\alpha,\beta$.

In the next two sections we shall consider two issues: the first, is the determination of the fields of interest given the position and the velocity of the interface; the second is the determination of the extra condition that provides us with the equation of motion for the interface.

\section{The Gibbs energy}
Let us denote by $\varrho(t)$ the distance of the interface from the center of the sphere, and by $\dot\varrho(t)$ its derivative, both at time $t$. In principle, when initial condition for concentration are prescribed, the evolution of $\varrho$ determines uniquely all fields of interest. Yet, the problem appears too complicated to extract useful information through an analytical approach. 

On the other hand, forward and reverse transformation take place in a neighborhood of the transition chemical potential $\mu_0$ at which the two phases would be in equilibrium in the absence of elastic misfit. This fact suggests that we restrict ouselves to situations where the chemical potential $\mu$ differs modestly from $\mu_0$. Accordingly, we may assume that the Gibbs energy depend linearly on $\mu-\mu_0$ and take:
 \begin{equation}
   \label{eq:53}
  \widehat{\omega}_\phi(\mu,\bm{\varepsilon})=-c_\phi(\mu-\mu_0)+\widehat W_\phi(\bm\varepsilon)
\end{equation}
where the \emph{strain energy} $W_\phi$ depends on the particular phase. An immediate consequence of \eqref{eq:53}, which is crucial to our developments, is that:
\begin{itemize}
\item in the bulk --- that is, away from the interface --- concentration is \emph{constant}. Precisely, we have:\footnote{Eq. \eqref{eq:54} means that if $x\in\Omega_\alpha(t)$ then $c(x,t)=c_\alpha$; if, instead,  $x\in\Omega_\beta(t)$ then $c(x,t)=c_\beta$.}
\begin{equation}
  \label{eq:54}
   c(x,t)=c_\phi\quad \textrm{for }x\in\Omega_\phi(t).
\end{equation}
\end{itemize}
As a result, the conservation of diffusant reads ${\rm div}\mathbf h=0$ and hence, in view of \eqref{eq:6}, \emph{chemical potential is harmonic in the bulk}, that is to say, it solves the Laplace equation:
\begin{equation}\label{eq:7}
  \Delta\mu=0
\end{equation}
away from the interface.

On taking as reference the state in which the material is in the $\alpha$ phase and on assuming isotropic linearly elastic response, we are led to the following choice for the \emph{strain energy}:\footnote{Here $|\bm\varepsilon|$ denoes the norm of the strain tensor, while ${\rm tr}$ is the trace operator.}
\begin{equation}
  W_\phi(\bm\varepsilon)=\begin{cases}
G|{\bm\varepsilon}|^2+ \frac \lambda 2|\tr{\bm\varepsilon}|^2\qquad\textrm{if }\phi=\alpha\\
G|{\bm\varepsilon}-{\bm\varepsilon}_{0}|^2+ \frac \lambda 2|\tr({\bm\varepsilon} -{\bm\varepsilon}_{0})|^2\qquad\textrm{if }\phi=\beta,
\end{cases}
\end{equation}
Here, $G$ and $\lambda$ are the standard Lam\'e moduli, assumed for simplicity to be equal in either phase and
\begin{itemize}
\item $\bm\varepsilon_0$ is the \emph{misfit strain between $\alpha$ and $\beta$ phase}.\footnote{In general, the misfit strain would depend not only on phase, but also on concentration. In particular, a concentration gradient may give rise to a stress for a single-phase specimen. In the present case, however, concentration is prescribed in each phase because of \eqref{eq:54}.}
\end{itemize}Consistent with the isotropy assumption on the strain energy, we take the misfit strain to be proportional to the identity tensor $\mathbf I$:\footnote{In Cartesian components, \eqref{eq:8} reads $(\bm\varepsilon_0)_{ij}=\epsilon_0\delta_{ij}$.}
\begin{equation}\label{eq:8}
  \bm\varepsilon_0=\epsilon_0\mathbf I,
\end{equation}
through a constant $\epsilon_0$.
\section{The boundary-value problems governing displacement and chemical potential}
We now can summarize the boundary-value problems governing the primary fields when the position $\varrho(t)$ of the interface and its velocity $V(t)=\dot\varrho(t)$ are known. The boundary, value problem governing displacement $\mathbf u$ and stress $\bm\sigma$ is:
\begin{equation}
  \label{eq:12}
  \begin{alignedat}{5}
&{\rm div}{\bm{\sigma}}=\mathbf 0
&\quad\quad&\textrm{in }\Omega_\alpha(t)\cup\Omega_\beta(t),&
\\
&{\bm{\sigma}}=2G\bm{\varepsilon}({\mathbf u})+\lambda({\rm div}{\mathbf u})\mathbf I
&\quad\quad&\textrm{in }\Omega_\alpha(t),&
\\
&{\bm{\sigma}}=2G\bm{\varepsilon}({\mathbf u})+(\lambda({\rm div}{\mathbf u})(2G+3\lambda)\epsilon_0)\mathbf I,
&\quad\quad&\textrm{in }\Omega_\beta(t),&
\\
&[\![{\mathbf u}]\!]=\mathbf 0
&\quad\quad&\textrm{on }\mathcal S(t),&
\\
&[\![{\bm{\sigma}}]\!]\mathbf n=\mathbf 0
&\quad\quad&\textrm{on }\mathcal S(t),&
\\
&{\bm{\sigma}}\mathbf n=\mathbf 0
&\quad\quad&\textrm{on }\partial\Omega.&
\end{alignedat}
\end{equation}
The boundary-value problem governing chemical potential $\mu$ and flux of diffusant $\mathbf h$ is:
\begin{equation}
  \label{eq:1211}
  \begin{alignedat}{5}
&{\rm div}{\mathbf h}=\mathbf 0
&\quad\quad&\textrm{in }\Omega_\alpha(t)\cup\Omega_\beta(t),&
\\
&{\mathbf h}=-m_\alpha\nabla\mu 
&\quad\quad&\textrm{in }\Omega_\alpha(t),&
\\
&{\mathbf h}=-m_\beta\nabla\mu 
&\quad\quad&\textrm{in }\Omega_\beta(t),&
\\
&[\![{\mu}]\!]=\mu_0,
&\quad\quad&\textrm{on }\mathcal S(t),&
\\
&[\![{\mathbf h}]\!]\cdot\mathbf n=(c_\beta-c_\alpha)V(t),
&\quad\quad&\textrm{on }\mathcal S(t),&
\\
&\mu=\mu_R
&\quad\quad&\textrm{on }\partial\Omega.&
\end{alignedat}
\end{equation}
On account of the spherical-symmetry features of domain and data, one can look for solutions of the form:
\begin{equation}
\label{rs}
\bfu(x,t)=u(r,t)\mathbf n(x),\qquad \mu(x,t)=v(r,t).
\end{equation}
where $r=r(x)=|x|$ is the distance of point $x$ from the center of the sphere and $\mathbf n(x)=\nabla r(x)$ is 
the radial vector field of unit norm. For the sake of brevity, we omit the analytical expressions of $u$ and $v$. Indeed, as we shall see in the next section, the only information we actually need to determine the evolution of the interface consists of:
\begin{itemize}
\item the value of chemical potential \emph{at the phase interface}:
\begin{equation}
  \label{eq:47}
  \mu=\mu_R+\frac{c_\beta{-}c_\alpha}{m_\beta}\Big(\frac{\varrho}{R}-1\Big)\varrho V;
\end{equation}
\item the jump of strain energy at the phase interface:
  \begin{equation}
\label{eq:41}
  W_\beta(\bm\varepsilon)-W_\alpha(\bm\varepsilon)=-\frac 2 3  \frac {G (2 G+3 \lambda )}{(2 G+\lambda )^2} \left(2G-3 \lambda -8 G \Big(\frac \rho R\Big)^3\right)\epsilon_0^2;
  \end{equation}
\item the jump of the normal derivative of the displacement gradient at the interface:
  \begin{equation}\label{eq:52}
    [\![\nabla\mathbf u]\!]\mathbf n=\frac{2G+3\lambda}{2G+\lambda}\epsilon_0\mathbf n;
  \end{equation}
\item the value of the radial stress at the interface:
  \begin{equation}
\label{eq:56}
    \sigma_r(\varrho)=\frac 4 3 G\frac{2G+3\lambda}{2G+\lambda}\Big(1-\Big(\frac\varrho R\Big)^3\Big)\epsilon_0^2.
  \end{equation}
\end{itemize}
To keep the present contribution reasonably short, we omit the details of the calculations leading to the above set of equations. These details shall be provided in \cite{DudaT2014inpreparation}.
\section{Configurational balance}
We may summarize the discussion in the previous sections in the following statements:
\begin{itemize}
\item if the radius $\varrho(t)$ of the $\alpha$ phase, and the velocity $V(t)=\dot\varrho(t)$ of the phase front are known, the boundary-value problems \eqref{eq:12}--\eqref{eq:1211} can be solved explicitly to obtain the instantaneous values of the fields of interest in the bulk;
\item an additional condition is needed in order to prescribe the motion of the interface. 
\end{itemize}
When seeking the aforementioned condition, it is important to bear in mind that the phase interface is composed of different material points at different times; therefore, it does not possess intrinsic material identity. Thus, the evolution of such material structure cannot be described by standard Newtonian forces, whose power expenditure is associated to the motion of individual material points. 

In order to construct models that can describe the evolution of material structures, the point of view adopted in \cite{Gurti2000Configurational} has been proven to be expedient. The main idea is to introduce forces that expend work on the motion of these structures and are subject to their own balance equations. In modern continuum mechanics, these forces are called \emph{configurational} or \emph{material}. This concept has been succesfully used to model the evolution of material structures such as phase interfaces, cracks, in three dimensional as well as lower dimensional continua, such as beams \cite{Tomas2011}.

As extensively discussed in \cite{Gurti2000Configurational}, the law governing the motion of a sharp interface can be derived from a \emph{configurational balance} $[\![\mathbf C]\!]\bfn+\mathbf g^{\mathscr S}+\mathbf e^{\mathscr S}=0$ involving a \emph{configurational stress}:
\begin{equation}
  \label{eq:17}
  \mathbf C=\omega\mathbf I-\nabla\mathbf u^T\bm{\sigma},
\end{equation}
with $\mathbf g^{\mathscr S}$ and $\mathbf e^{\mathscr S}$, respectively, the internal and external configurational force. It is worth noticing that the configurational stress, as defined in \eqref{eq:17}, is intimately related to the tensorial quantity introduced by Eshelby in his investigations on the motion of defects in elastic solids \cite{Eshel2006Continuum}.

For a \emph{structureless and dissipationless interface}, both the internal and the external configurational forces vanish, and the configurational balance reduces to:
\begin{equation}
  \label{eq:10}
  [\![\mathbf C]\!]\bfn=\mathbf 0.
\end{equation}
On taking the dot product of both sides of \eqref{eq:10} with $\mathbf n$, and on
 accounting for the continuity of traction and chemical potential across the interface, we arrive at:
\begin{equation}
  \label{eq:11}
  -(c_\beta-c_\alpha)(\mu-\mu_0)+W_\beta(\bm\varepsilon)-W_\alpha(\bm\varepsilon)-[\![\nabla\mathbf u]\!]\mathbf n\cdot\bfS\mathbf n=0.
\end{equation}
On substituting \eqref{eq:47}--\eqref{eq:56} into \eqref{eq:11} we obtain one of the main result of this paper, namely, the evolution equation for $\varrho(t)$, the radius of the $\alpha$ phase:
\begin{multline}
  \label{eq:50}
  \Big(\frac{\varrho}{R}-1\Big)\frac{c_\beta{-}c_\alpha}{m_\beta}\varrho\dot\varrho\\
=(c_\beta{-}c_\alpha)\mu_0\left(1-\frac{\mu_R}{\mu_0}\right)+2 \epsilon_0^2 G \frac{2 G+3 \lambda}{2 G+\lambda} \left(2 \left(\frac{\varrho}{R}\right)^3-1\right).
\end{multline}
\section{Transformation time}
With a view towards extracting information from \eqref{eq:50}, we introduce the \emph{characteristic time}:
\begin{align}
  \tau=\frac{R^2}{m_\beta \mu_0},
\end{align}
and we introduce the dimensionless quantities:
\begin{equation}\label{eq:59}
y=\frac \varrho R,\qquad a=\frac{\mu_R}{\mu_0},\qquad  b=\frac{2G}{2G+\lambda}
\frac{(2G+3\lambda)\epsilon_0^2}
{\mu_0(c_\beta-c_\alpha)}.    
\end{equation}
Then, \eqref{eq:50}  takes the following form:
\begin{align}\label{eq:25}
 \tau y(1-y)\dot y=1-a+b(2y^3-1).
\end{align}
We now look for solutions of \eqref{eq:59} such that $\varrho(0)=1$ and $\dot\varrho<0$. Any such solution describes a physical process in which the specimen is initially in the $\alpha$ phase, the $\beta$ phase nucleates at the boundary, and the phase interface moves towards the interior of the particle. This process is accompanied by adsorption from the reservoir until the specimen is saturated, a situation that corresponds to $\varrho=0$, when the particle is entirely in the $\beta$ phase. We refer to this process as the $\alpha\to\beta$ transformation. It turns out that:
\begin{itemize}
\item the $\alpha\rightarrow\beta$ transformation can take place only if the following condition is satisfied:
  \begin{equation}
   \label{eq:1} 
a>1+b;
  \end{equation}
moreover, if \eqref{eq:1} holds, then the transformation time is:
\begin{align}\label{eq:4}
    T_{\alpha\rightarrow \beta}=\frac \tau{a-1} f(\kappa),
\end{align}
where the parameter $\kappa$ and the function $f$ are defined by, respectively,
$$\kappa=\frac b {1-a}\qquad \textrm{and}\qquad f(\kappa)=\int_0^1 \frac{z(1-z)}{1+\kappa(2z^3-1)}{\rm d}z.$$
\end{itemize}
In a similar fashion, we can consider a solution such that $\varrho(0)=0$ and $\dot\varrho>0$. This solution describes a physical process in which the specimen is initially in the the $\beta$ phase, the $\alpha$ phase nucleates at the center, and the phase front moves towards the boundary until the specimen is completely in the $\alpha$ phase. It turns out that: 
\begin{itemize}
\item the $\beta\rightarrow\alpha$ transformation can take place only if $a<1-b$; in this case, the transformation time is:
\begin{align}\label{eq:2}
    T_{\beta\rightarrow \alpha}=\frac \tau{1-a} f(\kappa).
\end{align}
\end{itemize}
It is worth noticing that $f$ has a closed-form expression, as can be verified using standar computer algebra systems. The graph of $f$ is plotted in Figure 2. Positive values of $\kappa$ are relevant in the $\alpha\to\beta$ transformation. Negative values are for the $\beta\to\alpha$ transformation. 
\begin{figure}[h]
\begin{center}
\footnotesize
\def\svgwidth{0.73\linewidth}
\qquad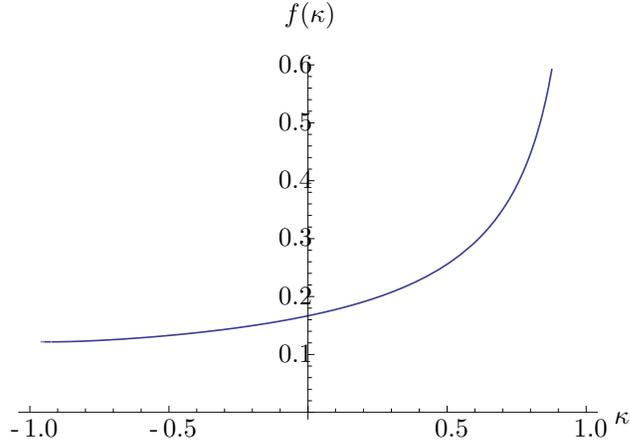
\footnotesize\caption{Plot of $f(\kappa)$. Positive values of $\kappa$ yield the transition time for the $\beta\to\alpha$ transformation. Negative values of $\kappa$ yield the transformation time for the $\alpha\to\beta$ transformation. Remarkably, elastic misfit slows down the $\beta\to\alpha$ transformation and speeds up the $\alpha\to\beta$ transformation.}
\end{center}  
\end{figure}
From \eqref{eq:4} and \eqref{eq:2} we can draw the following important consequence:
\begin{itemize}
\item the ratio between the transition time with and without misfit strain is given by $f(\kappa)/f(0)$.
\end{itemize}
On taking into account the above observation, and on looking at the plot of $f$ for $\kappa\in[0,1)$, we reckon that the elastic misfit between the two phases substantially slows down the transition from the $\beta$ to the $\alpha$ phase, with the tranformation time blowing up as $\kappa\to 1$. Concerning the $\alpha\to\beta$ transformation, we observe, on the other hand, that an increase of the elastic misfit yields a decrease of $\kappa$ and in turn, as is apparent from Figure 2, a reduction of the transformation time. We therefore conclude that the presence of a lattice misfit accelerates the transformation from the $\alpha$ to the $\beta$ phase and plays a crucial role in the kinetics.

As a final remark, we observe that in order to trigger the $\alpha\to\beta$ transformation and the $\beta\to\alpha$ transformation the chemical potential must satisfy, respectively, $\mu_R>\mu_R^{\alpha\to\beta}:={\mu_0(1+b)}$ and $\mu_R<\mu_R^{\beta\to\alpha}:={\mu_0(1-b)}$. In particular, if the chemical potential of the reservoir oscillates between $\mu_R^{\rm \alpha\to\beta}$ and $\mu_R^{\beta\to\alpha}$, then the path of $y$ versus $\mu_R$  would define an hysteresis loop, consistent with the results in \cite{SchwaK1995Thermodynamics}.

\section{Conclusions}
We have considered
the continuum theory proposed in \cite{fried1999coherent} describing diffusion of a chemical species in an elastic solid partitioned in two phases separated by a sharp interface.
We have specialized the theory to a spherical domain containing a
structureless and dissipationless concentric phase interface,
and we have investigated the effects of phase-trasformation strain on the kinetic of the interface.

A key point of our approach is the assumption that the Gibbs energy depend linearly on the difference between the local value of chemical potential and a reference value. The main consequence of this assumption is that concentration is constant in each phase. This state of matters makes the ensuing mathematical problem tractable, for it allows us to compute explicitly the instantaneous values of displacement and chemical potential in terms of the position of the interface. When this information is fed to the configurational balance equation, we obtain an autonomous first-order differential equation governing the position of the interface. Our analytical treatment of this equation shows that the presence of a lattice misfit between the two phases accelerates phase transformation processes which start with one phase nucleating at the boundary, and slows down processes that begin with nucleation of one phase from the center. Furthermore, our analysis confirms that elastic mistfits, besides being responsible for hysteresis in the phase-transformation process, strongly affect its  kinetics. All these effects can be quantified in terms of a modest number of parameters.

\section{Acknowledgments}
The first author gratefully acknowledges financial support provided by  CNPq (312153/2013-9). The second author acknowledges financial support by the Italian INdAM GNFM through initiative ``Progetto Giovani''.

\section*{\large References}
\bibliographystyle{elsarticle-num}
\bibliography{bibliography_hydrogen_Rio,bibliography}

\end{document}

%% file: sphere.pdf_tex
\begingroup%
  \makeatletter%
  \providecommand\color[2][]{%
    \errmessage{(Inkscape) Color is used for the text in Inkscape, but the package 'color.sty' is not loaded}%
    \renewcommand\color[2][]{}%
  }%
  \providecommand\transparent[1]{%
    \errmessage{(Inkscape) Transparency is used (non-zero) for the text in Inkscape, but the package 'transparent.sty' is not loaded}%
    \renewcommand\transparent[1]{}%
  }%
  \providecommand\rotatebox[2]{#2}%
  \ifx\svgwidth\undefined%
    \setlength{\unitlength}{277.20737305bp}%
    \ifx\svgscale\undefined%
      \relax%
    \else%
      \setlength{\unitlength}{\unitlength * \real{\svgscale}}%
    \fi%
  \else%
    \setlength{\unitlength}{\svgwidth}%
  \fi%
  \global\let\svgwidth\undefined%
  \global\let\svgscale\undefined%
  \makeatother%
  \begin{picture}(1,0.98410081)%
    \put(0,0){\includegraphics[width=\unitlength]{sphere.pdf}}%
    \put(0.27354454,0.19106208){\color[rgb]{0,0,0}\makebox(0,0)[lb]{\smash{$R$}}}%
    \put(0.38792059,0.57035528){\color[rgb]{0,0,0}\makebox(0,0)[lb]{\smash{$\rho$}}}%
    \put(0.49414562,0.39653046){\color[rgb]{0,0,0}\makebox(0,0)[lb]{\smash{$\Omega_\alpha$}}}%
    \put(0.34960149,0.79828761){\color[rgb]{0,0,0}\makebox(0,0)[lb]{\smash{$\Omega_\beta$}}}%
  \end{picture}%
\endgroup%

%% file: fk.pdf_tex
\begingroup%
  \makeatletter%
  \providecommand\color[2][]{%
    \errmessage{(Inkscape) Color is used for the text in Inkscape, but the package 'color.sty' is not loaded}%
    \renewcommand\color[2][]{}%
  }%
  \providecommand\transparent[1]{%
    \errmessage{(Inkscape) Transparency is used (non-zero) for the text in Inkscape, but the package 'transparent.sty' is not loaded}%
    \renewcommand\transparent[1]{}%
  }%
  \providecommand\rotatebox[2]{#2}%
  \ifx\svgwidth\undefined%
    \setlength{\unitlength}{367.98959941bp}%
    \ifx\svgscale\undefined%
      \relax%
    \else%
      \setlength{\unitlength}{\unitlength * \real{\svgscale}}%
    \fi%
  \else%
    \setlength{\unitlength}{\svgwidth}%
  \fi%
  \global\let\svgwidth\undefined%
  \global\let\svgscale\undefined%
  \makeatother%
  \begin{picture}(1,0.58445442)%
    \put(0,0){\includegraphics[width=\unitlength]{fk.pdf}}%
    \put(-0.00106151,0.00030436){\color[rgb]{0,0,0}\makebox(0,0)[lb]{\smash{-}}}%
    \put(0.01633028,0.00030436){\color[rgb]{0,0,0}\makebox(0,0)[lb]{\smash{1.0}}}%
    \put(0.18372632,0.00030436){\color[rgb]{0,0,0}\makebox(0,0)[lb]{\smash{-}}}%
    \put(0.20111812,0.00030436){\color[rgb]{0,0,0}\makebox(0,0)[lb]{\smash{0.5}}}%
    \put(0.56199788,0.00030436){\color[rgb]{0,0,0}\makebox(0,0)[lb]{\smash{0.5}}}%
    \put(0.74678571,0.00030436){\color[rgb]{0,0,0}\makebox(0,0)[lb]{\smash{1.0}}}%
    \put(0.35438332,0.10012426){\color[rgb]{0,0,0}\makebox(0,0)[lb]{\smash{0.1}}}%
    \put(0.35438332,0.17711744){\color[rgb]{0,0,0}\makebox(0,0)[lb]{\smash{0.2}}}%
    \put(0.35438332,0.25411059){\color[rgb]{0,0,0}\makebox(0,0)[lb]{\smash{0.3}}}%
    \put(0.35438332,0.33110378){\color[rgb]{0,0,0}\makebox(0,0)[lb]{\smash{0.4}}}%
    \put(0.35438332,0.40809695){\color[rgb]{0,0,0}\makebox(0,0)[lb]{\smash{0.5}}}%
    \put(0.35438332,0.48509012){\color[rgb]{0,0,0}\makebox(0,0)[lb]{\smash{0.6}}}%
    \put(0.36179376,0.5514202){\color[rgb]{0,0,0}\makebox(0,0)[lb]{\smash{$f(\kappa)$}}}%
    \put(0.80058455,0.01846214){\color[rgb]{0,0,0}\makebox(0,0)[lb]{\smash{$\kappa$}}}%
  \end{picture}%
\endgroup%